\documentclass{cimento}
\usepackage{graphicx} 
\title{Microlensing in Galactic Halos}
\author{S.~Calchi Novati\from{ins:a}}
\instlist{\inst{ins:a} Dipartimento di Fisica ``E. R. Caianiello'', Universit\`a
di Salerno, 84081 Baronissi (SA), and INFN, sez. Napoli, Italy}
\PACSes{\PACSit{95.75.De},{98.35.Gi},{95.35.+d}}

\begin{document}

\maketitle

\begin{abstract}
In the framework of the search of dark matter
in galactic halos in form of massive compact halo object (MACHOs),
we discuss the status of microlensing observations
towards the Magellanic Clouds and the Andromeda galaxy, M31.
The detection of a few microlensing events has been reported,
but an unambiguous conclusion on the halo content in form
on MACHOs has not been reached yet. A more detailed modelling
of the expected signal and a larger statistics of observed
events are mandatory in order to shed light on this important
astrophysical issue. 
\end{abstract}

\section{Introduction}

Gravitational microlensing, as first noted in \cite{ref:pacz86}, is a very efficient
tool for the detection and the characterisation 
of massive astrophysical halo compact objects (MACHOs),
a possible component of dark matter halos. 
Following the first exciting detection of  microlensing
events \cite{ref:macho93,ref:eros93,ref:ogle93},
by now the detection of $\sim 30$ events have been reported
towards the Magellanic Clouds and our nearby galaxy, M31,
and first interesting conclusions on this issue have been reported
(Section~\ref{sec:LMC} and Section~\ref{sec:M31}).
Soon enough, however, the Galactic bulge probed to be
an almost an interesting target \cite{ref:pacz91},
and indeed by now the number of observed microlensing
events along this line of sight exceeds by two order of magnitudes 
that observed towards the Magellanic Clouds and M31.
In that case the contribution from the dark matter
halo is expected to be extremely small compared to that
of either bulge or disc (faint) stars \cite{ref:griest91}. Microlensing
searches towards the Galactic bulge are therefore important
as they allow to constrain the inner Galactic structure \cite{ref:pacz94}.
Recently, the MACHO \cite{ref:popowski05}, OGLE \cite{ref:sumi06} and 
EROS \cite{ref:hamadache06} collaborations presented the results
of their observational campaigns towards this target.
A remarkable conclusion is the agreement among these different
searches as for the observed value of the optical depth
and the agreement with the theoretical expectations 
\cite{ref:evans02,ref:hangould03}. 
For a more recent  discussion see also \cite{ref:novati07}, 
where the issue of the bulge mass spectrum is treated.

\section{The microlensing quantities}
\label{sec:ml}

Microlensing events are due to a lensing object
passing near the line of sight towards a background
star. Because of the event configuration, the observable 
effect during a microlensing event
is an apparent transient amplification
of the star's flux (for a review see e.g. \cite{ref:roulet97}).  

The \emph{optical depth} is the instantaneous
probability that at a given time a given star is amplified
so to give rise to an observable event. This quantity
is the probability to find a lens within the ``microlensing tube'',
a tube around the line of sight of (variable) radius equal to the \emph{Einstein radius}, 
$R_\mathrm{E}=\sqrt{4G\mu_l/c^2\, D_l D_{ls}/D_s}$, where $\mu_l$
is the lens mass, $D_l,\,D_s$  are the distance to
the lens and to the source,  respectively, and $D_{ls}=D_s-D_l$.
The optical depth reads
\begin{equation}
\label{eq:tau}
\tau = \frac{4\pi G D_s^2}{c^2}\int_{0}^{D_s} \mathrm{d}x \rho(x) x(1-x)\,,
\end{equation}
where $\rho$ is the \emph{mass} density distribution of lenses
and $x\equiv D_l/D_s$. The optical depth provides valuable informations on the overall 
density distribution of the lensing objects,
but it can not be used to further characterise
the events, in particular, it does  not depend on the lens
mass. This is because lighter (heavier) objects are, 
for a given total mass of the lens population,
more (less) numerous but their lensing cross section is smaller (larger),
and the two effects cancel out. 
The optical depth turns out to be an extremely small quantity,
of order of magnitude $\sim 10^{-6}$. This  implies that
one has to monitor extremely large sets of stars
to achieve a reasonable statistics. 

The experiments measure the number of the events
and their characteristics, in particular their durations.
To evaluate these quantities one makes use of the
microlensing \emph{rate} that expresses the number of lenses
that pass through the volume element of the microlensing tube $\mathrm{d}^3x$
in the time interval $\mathrm{d}t$ for a given lens number density distribution
$n(\vec{x})$ and velocity distribution $f(\vec{v})$
\begin{equation}
\label{eq:rate}
\mathrm{d} \Gamma = \frac{n_l\,\mathrm{d}^3 x}{\mathrm{d}t}
\times f(\vec{v}_l) \mathrm{d}^3 v_l\,.
\end{equation}
The volume element of the microlensing tube is 
$\mathrm{d}^3 x=(\vec{v}_{r\bot} \cdot \hat{\vec{n}}) \mathrm{d}t \mathrm{d}S$.
$\mathrm{d}S=\mathrm{d}l\mathrm{d}D_l$  is the portion of the tube external surface,
and $\mathrm{d}l=u_t R_\mathrm{E} \mathrm{d}\alpha$, 
where $u_t$ is the maximum impact parameter, 
$\vec{v}_{r}$ is the lens relative velocity with respect
to the microlensing tube
and $\vec{v}_{r\bot}$ its component in the plane
orthogonal to the line of sight, 
and $\hat{\vec{n}}$ is the unit vector normal to the tube inner surface 
at the point where  the microlensing tube is crossed by the lens.
The velocity of the lenses entering the tube is $\vec{v}_l=\vec{v}_r+\vec{v}_t$.
$\vec{v}_t$ is the tube velocity.

The differential rate is directly related to the number
of expected microlensing events as 
$\mathrm{d}N=N_\mathrm{obs} T_\mathrm{obs} \mathrm{d}\Gamma$,
where $N_\mathrm{obs}, T_\mathrm{obs}$ are the number of monitored sources
and the whole observation time, respectively.
Furthermore, the distribution for 
the duration of the microlensing events, the \emph{Einstein time},
$t_\mathrm{E}=R_\mathrm{E}/v_{r\bot}$, can also be deduced from
the differential microlensing rate, as $\mathrm{d}\Gamma/\mathrm{d}t_\mathrm{E}$.
Besides on the lens mass, the key quantity one is usually interested into,
$t_\mathrm{E}$ depends also on other usually unobservable quantities.
It is therefore suitable to observe a large enough
number of events so to be able to deal 
statistically with the degeneracies intrinsic
to the parameter space of microlensing events.

Eventually note that, in calculating the microlensing quantities, the optical depth
and the rate, one can also take into account
the source spatial and velocity distributions. 

\section{Microlensing towards the Magellanic Clouds}
\label{sec:LMC}

\begin{figure}
\includegraphics[width=7cm]{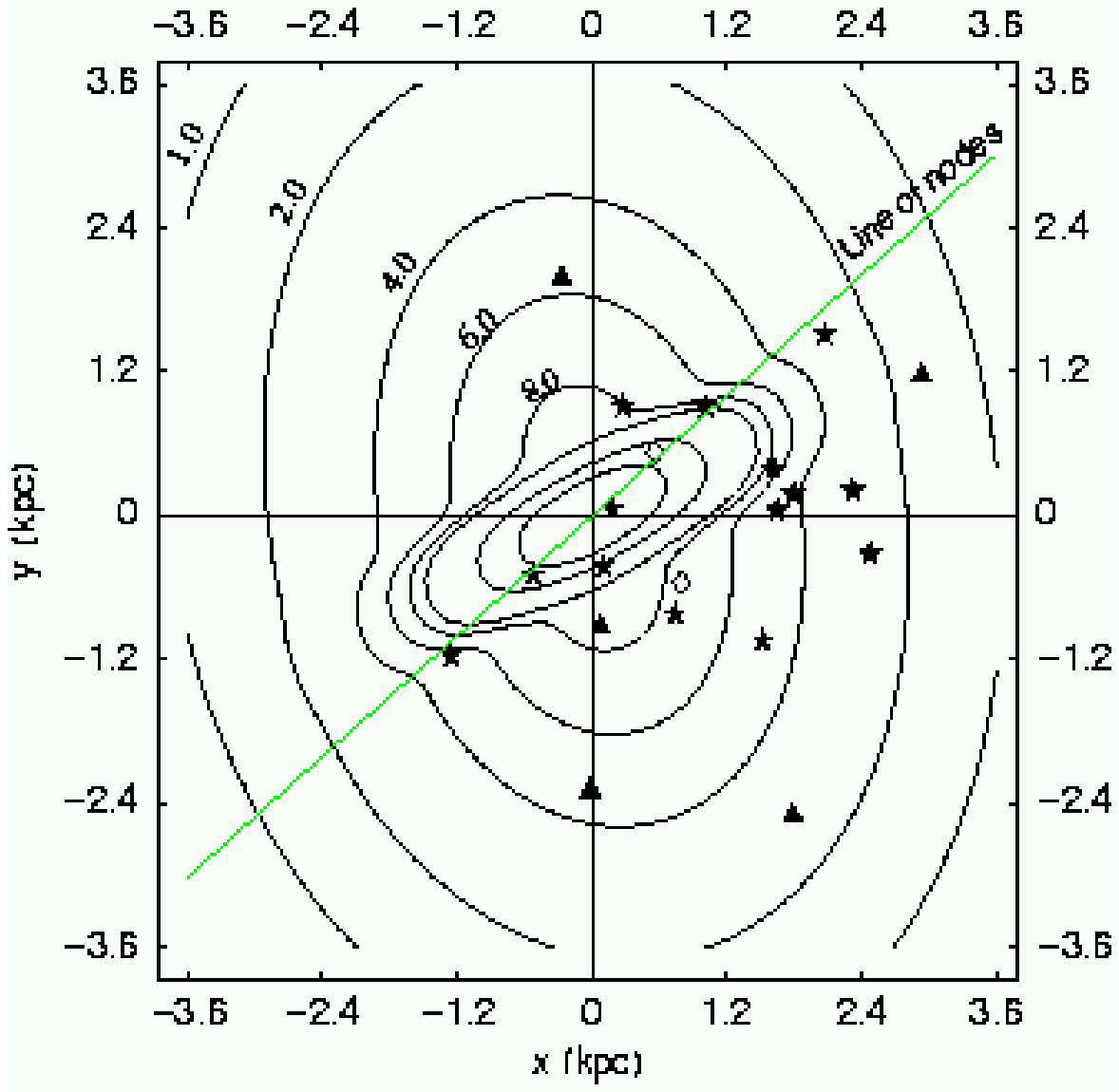}    
\includegraphics[width=7cm]{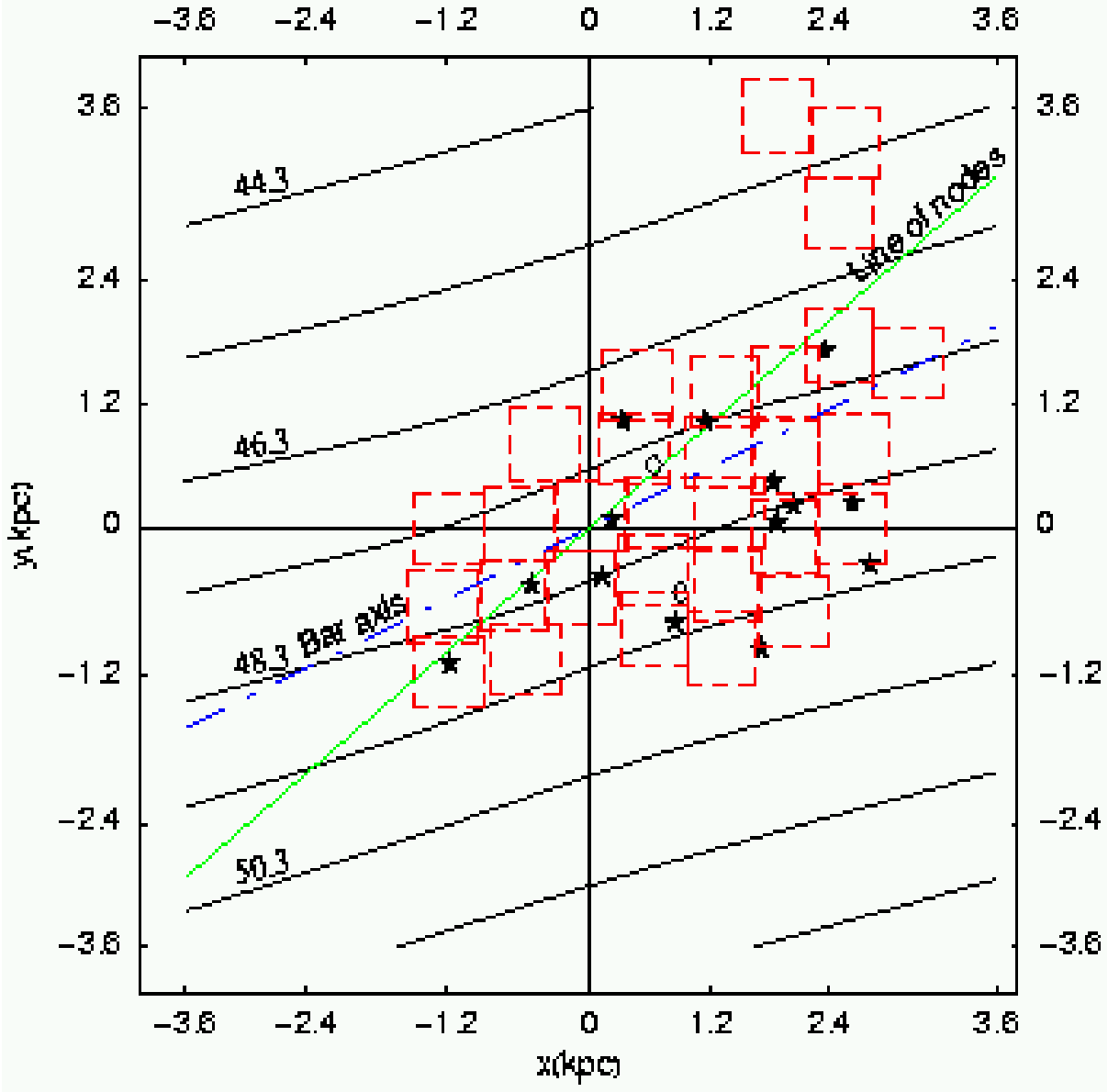}
\includegraphics[width=7cm]{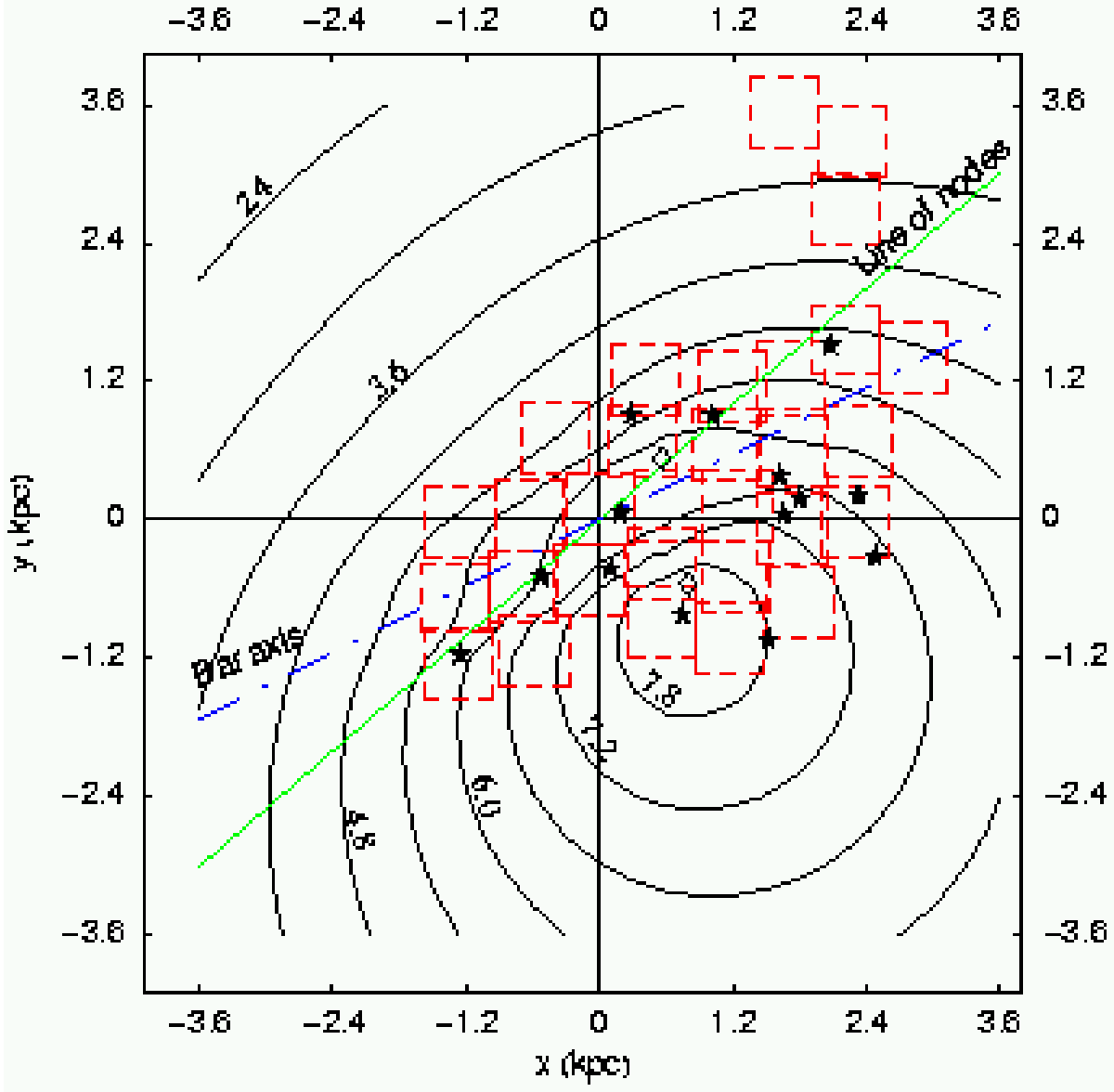}
\includegraphics[width=7cm]{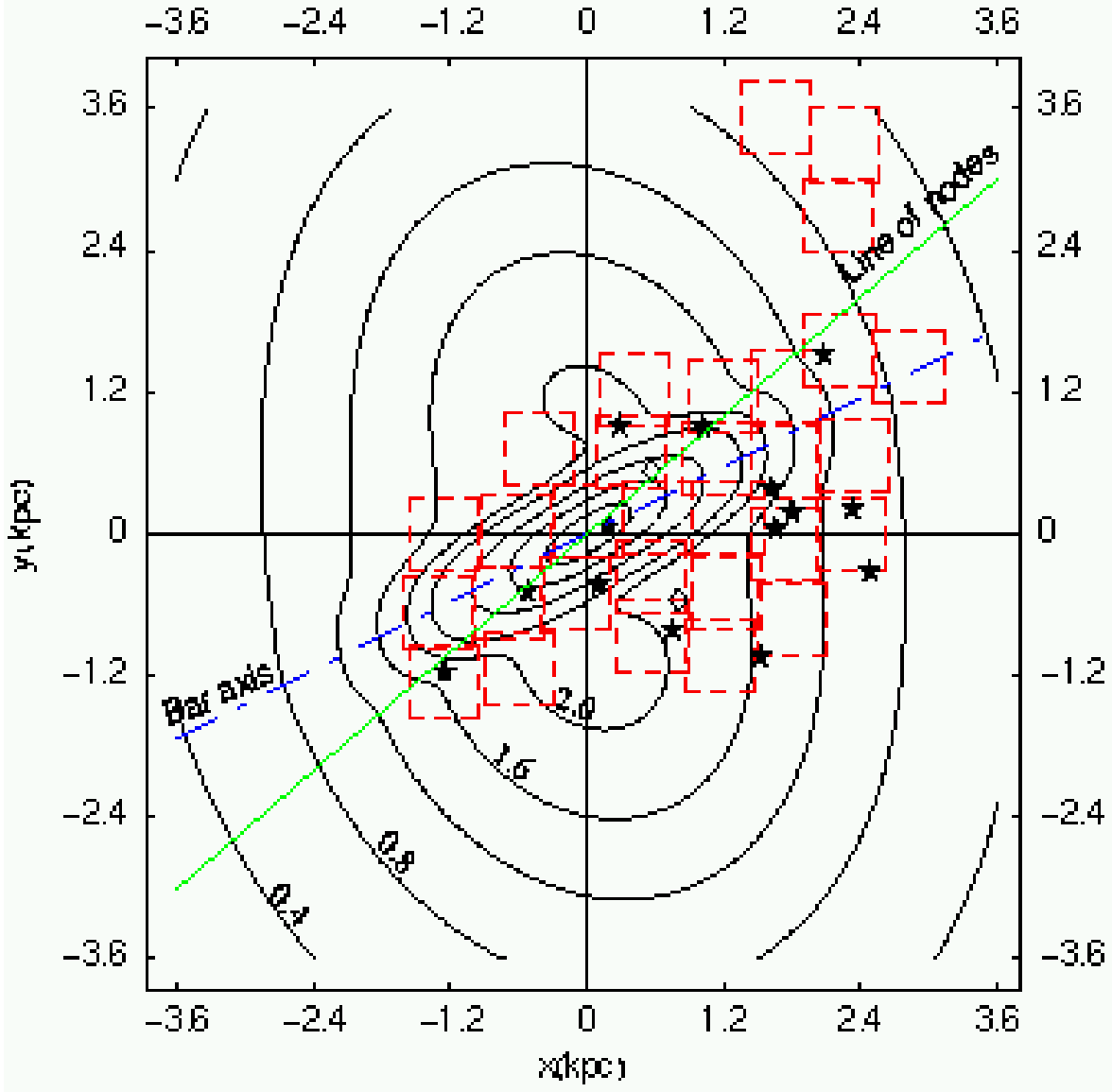}
\caption{Top left: projection on the sky plane of the column density
of the LMC disc and bar. The numerical values on the contours are in
$10^7~\mathrm{M}_\odot~\mathrm{kpc}^{-2}$ units. The three innermost contours
correspond to 10, 20 and $30\times 10^7~\mathrm{M}_\odot~\mathrm{kpc}^{-2}$.
The location of the MACHO (black stars and empty diamonds) and EROS (triangles)
microlensing candidates are shown. The $x-y$ axes are directed towards West and North
respectively. From top right to bottom left: contours maps of the optical depth
for lenses in the Galactic halo, LMC halo and self lensing, respectively.
The numerical values are in $10^{-8}$ units. Also shown, the location
of the fields observed by the MACHO collaboration. (Figures adapted
from \cite{ref:mancini04}.)}
\label{fig:lmc-tau}
\end{figure}

The first survey aimed at the detection of 
microlensing events have been carried out
towards the Large and Small Magellanic Clouds 
(LMC and SMC, respectively), so to probe the  MACHO content 
within the Galactic halo. The main results
have been obtained by the MACHO \cite{ref:macho00}
and the EROS \cite{ref:eros07} collaborations.

MACHO reported the detection of 13-17 microlensing
events towards the LMC, and concluded that a rather significant
(mass) fraction of the Galactic halo, $f\sim~20\%$,
is made up of dark mass objects of $\sim~0.4~\textrm{M}_\odot$.
On the other hand, EROS reported the detection
of 1 event towards the SMC and no events
towards the LMC, whereas they evaluated, for a full halo 
of $0.4~\textrm{M}_\odot$ MACHO, an expected number of microlensing events 
$\sim 39$. Correspondingly, EROS put a rather severe \emph{upper} limit
on the halo fraction in form of MACHOs, $f<0.08$ for $0.4~\textrm{M}_\odot$ MACHO objects.

The disagreement between the results obtained by
the MACHO and the EROS collaboration
leaves the issue of the halo content 
in form of MACHOs open. A possible issue 
is the  nature of the flux
variations reported by the MACHO collaboration.
Indeed, microlensing searches are plagued by 
variable stars (that represent the overwhelmingly
majority of the flux variations detected) 
masquerading as microlensing events.
However, Bennet \cite{ref:bennet05} performed 
a new analysis on the MACHO data set concluding
that ``[\ldots] the main conclusions of the MACHO LMC
analysis are unchanged by the variable star contamination[\ldots]''.

Looking for MACHO events, the second
background is constituted by
``self-lensing'' events, where, besides the source,
also the lens belongs to some luminous
star population (either in the LMC itself
or possibly, along the line of sight, in the Galactic disc).
This possibility was first addressed  in \cite{ref:sahu94,ref:wu94}
and has been further discussed by several authors  
(e.g. \cite{ref:gould95,ref:gyuk00}).

Besides these possible background contaminations,
a few aspects of the EROS analysis are worth
being mentioned. First, while the fields
observed by MACHO towards the LMC are all concentrated
around the central region, EROS monitored an extremely
larger region. This  alleviates the issue of self lensing
but also that of a possible clumpiness of the Galactic halo
right along the line of sight towards the LMC (this argument
is balanced, however, by the much smaller expected rate
in the outer with respect to the inner LMC regions).
Second, EROS restricted his analysis
to a subsample of bright sources, this choice
being motivated by the superior photometric precision
of the corresponding light curves (so to reduce
possible contamination from variable stars),
and by the possibility of a better understanding
of the so-called ``blending'' effect. 
The latter issue is of particular relevance in microlensing analyses
and it concerns the ability to correctly evaluate the source flux,
in absence of amplification, in crowded fields where the observed objects
can be, to some extent, the blend of several stars.
It is worth stressing that a similar approach was crucial
to get to the agreement between the 
theoretical expectations and the measured values of the optical
depth in the case of observations towards the Galactic bulge.

\begin{figure}
\includegraphics[width=11cm]{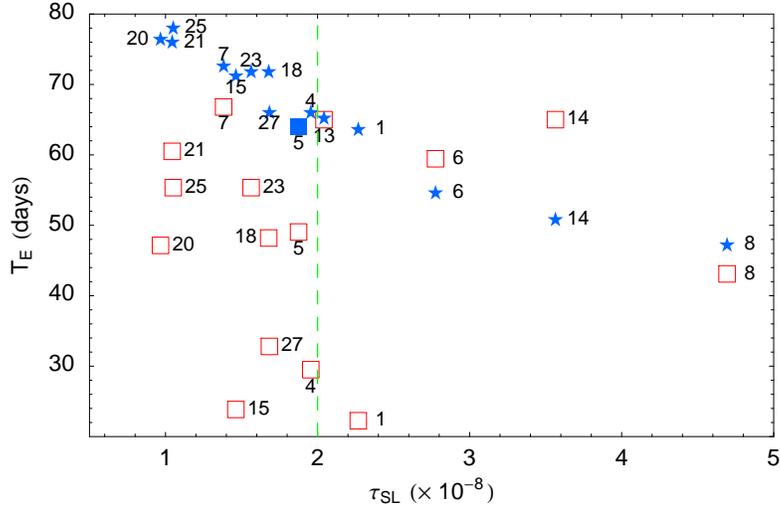}    
\caption{Scatter plot of the observed values (empty boxes) of the Einstein time
and of the expected values of the median duration (filled stars) with respect
to the self-lensing optical depth evaluated along the direction of the events.
The dashed line for $\tau_\mathrm{SL}=2$ approximately delimits the inner LMC region,
where a good agreement is found between the two values for most of the observed events,
and the outer region, where the rise in the expected duration is clearly
not observed. (Figure adapted from \cite{ref:mancini04}.)}
\label{fig:lmc-rate}
\end{figure}

\begin{figure}
\includegraphics[width=9cm]{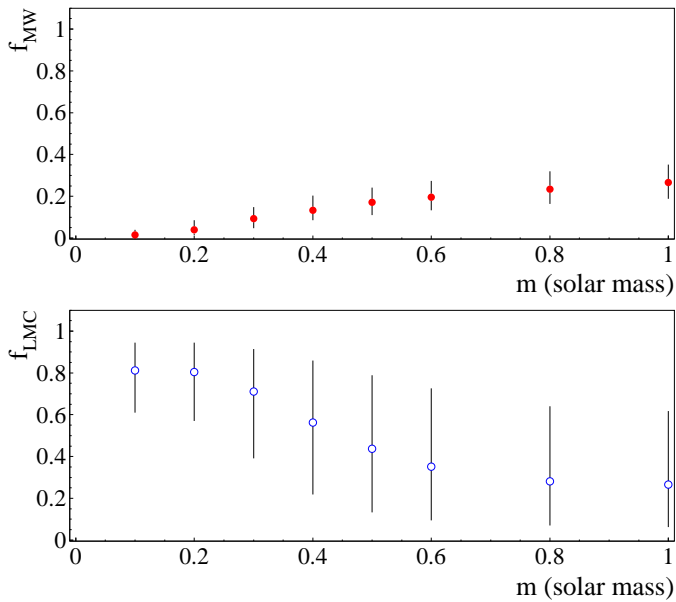}    
\caption{Galactic (top) and LMC dark matter halo fraction, median value
with 68\% CL error, as a function of the MACHO mass. For values in the mass range
$(0.1-0.3)~\mathrm{M}_\odot$, preferred for LMC lenses on the basis of an analysis
of the event duration and spatial distributions, the LMC halo dark matter fraction
turns out to be significantly larger than the Galactic one. 
(Figure adapted from \cite{ref:novati06}.)
}
\label{fig:lmc-halo}
\end{figure}

More recently, new analyses of the MACHO results have been undertaken.
In \cite{ref:jetzer02} it is shown that the observed events
are probably distributed among different components
(disc, Galactic halo, the LMC halo and self lensing).
Taking advantage of a new modelling of both the luminous
and the dark components of the LMC \cite{ref:vdm02},
in \cite{ref:mancini04} the self-lensing  issue has been once
more addressed considering  the set of microlensing events
reported by the MACHO collaboration. In Figure~\ref{fig:lmc-tau} the density profile
of the luminous components of the LMC, disc and bar, is shown
together with the optical depth profiles for the Galactic halo,
the LMC halo and self lensing. 
Furthermore, through an analysis of the differential microlensing
rate it has been shown that self-lensing events cannot contribute
to all of the $\sim~10$ observed events. First, the expected number
for self lensing turns out to be significantly smaller, about 1-2 events
at most. Second (Figure~\ref{fig:lmc-rate}), 
for self-lensing events one expects a peculiar 
signature in the relationship between the event duration
and their spatial distribution, with longer durations
expected in the outer LMC region (where, correspondingly,
the self-lensing optical depth turns out to be smaller). Such 
a relationship, however, is not observed. 
The same set of MACHO events
has also been analysed in \cite{ref:novati06}, where in particular
the question of a possible significant contribution to the observed events
of lenses belonging to the dark matter halo of the LMC, as opposed to those
of the Galactic halo population, has been addressed
(this possibility had previously been discussed in \cite{ref:gould93,ref:kerins99}). 
In particular, studying 
both the spatial and the duration distributions,
it is shown that only a fraction of the events have characteristics
that match those expected for the latter population, hinting
that a population of somewhat lighter, $\sim~0.2~\mathrm{M}_\odot$,
LMC halo MACHO may indeed contribute to the observed events.
Challenging the usual assumption of equal halo fractions
in form of MACHO for the Galactic and the LMC halo it was then shown,
Figure~\ref{fig:lmc-halo}, 
that indeed for MACHO masses in the range $(0.1-0.3)~\mathrm{M}_\odot$
the LMC halo mass fraction can be significantly larger than the Milky Way's
so that up to about half of the observed events could indeed be attributed
to the LMC MACHO dark matter halo.

\section{Microlensing towards M31}
\label{sec:M31}

The contradictory results of the microlensing campaigns towards
the Magellanic Clouds challenge to probe the MACHO distribution
along different line of sights. The Andromeda galaxy, M31, 
nearby and similar to the Milky Way, is a suitable target for this search
\cite{ref:crotts92,ref:baillon93,ref:jetzer94}. First,
it allows us to explore the Galactic halo along a different line of sight.
Second, it has its own dark matter halo that, as we look at it from outside, 
can be studied globally. We stress that this is a fundamental advantage with respect
to the Magellanic Clouds searches. The analysis shows that,
as an order of magnitude, for a given MACHO mass and halo fraction,
one expects about 3 microlensing events
by MACHOs in the M31 halo for each event due to a MACHO in the Galactic halo
(in fact, in the latter case the number of available lenses is enormously
smaller, by about 4 orders of magnitude, but this fact is, almost,
balanced by the much larger value of the Einstein radius). 
Eventually, the high inclination
of the M31 disc is expected to provide a strong gradient in the spatial
distribution of microlensing events, that can in principle give an unmistakable
signature for M31 microlensing halo events. 

\begin{figure}
\includegraphics[width=8cm]{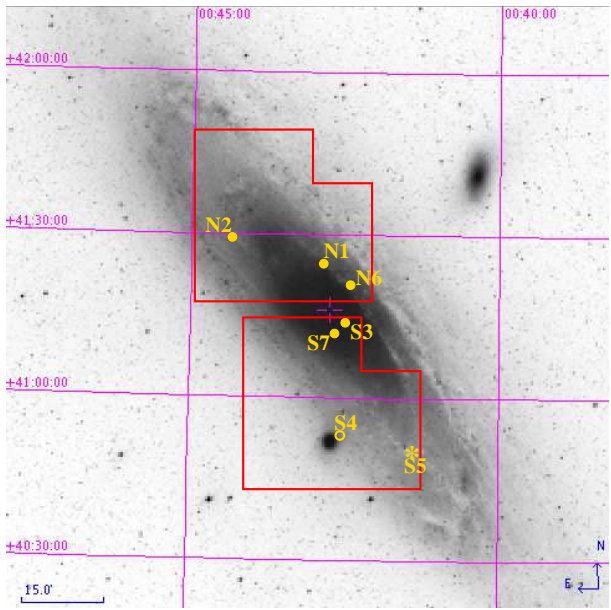}    
\caption{Projected on M31, the boundaries of the observed INT fields
are shown together with the position of the microlensing candidates (circles)
reported by the POINT-AGAPE collaboration. 
Note in particular the position of the microlensing event N2,
located rather far away from the M31 centre, where the expected self-lensing signal
is very low. Note also that S4, the M31-M32 microlensing event,
and S5, a possible binary event not included in the selection,
are not included in the analysis for the determination of the halo fraction
in form of MACHOs. (Figure adapted from \cite{ref:novati05}.)}
\label{fig:int-fields}
\end{figure}

As compared to that to the LMC and the SMC, $\sim 50~\mathrm{kpc}$,
the distance to M31 is, by more than an order of magnitude, larger,
$\sim 770~\mathrm{kpc}$. As a consequence, the potential sources
of microlensing events are not going to be, as for the Magellanic Clouds,
resolved objects. This calls for a peculiar technique, usually referred to as
``pixel-lensing'', whose key feature is the fact that one monitors
flux variations of unresolved objects 
in each pixel element of the image \cite{ref:gould96}.
Although, in principle, \emph{all} stars in the pixel field
are possible sources one can only detect lensing events
due either to bright enough stars or to extremely high amplification events
(in any case all stars in the pixel field contribute
to the overall flux background).
As it has been first shown by the AGAPE group \cite{ref:agape97},
the former case is by far the most likely, with, in any case, a 
number of potential sources
per arsec-square that can easily exceed, in the more crowded region,
a few hundreds. With respect to the analysis towards the Magellanic Clouds
this means that a huge number of sources is potentially available.
However, one must handle with the difficulty that the source flux
is not a directly observable quantity. This adds a further degeneracy
in the microlensing parameter space, in particular it does not allow
one to unambiguously determine, out of the observed event duration,
the Einstein time. Instead, what is directly observable is the so-called
$t_{1/2}$, the full-width-at-half-maximum of the flux variation
visible above the background. It turns out that, for self-lensing events
as well for MACHO events in the mass range preferred  by the Magellanic Clouds searches,
$t_{1/2}$ is of order of only a few days. This is shorter than the typical durations
observed towards the Magellanic Clouds, and this is relevant in the analysis as it 
makes easier the distinction between the contaminating background 
of variable stars and the truly microlensing flux variations.

The first convincing detection of a microlensing event towards M31
has been reported by the AGAPE  group \cite{ref:agape99}. At the same time,
other collaborations
have undertaken searches for microlensing towards M31 and the detection
of a few more candidate events has been discussed: 
Columbia-VATT \cite{ref:crotts96}, POINT-AGAPE \cite{ref:auriere01,ref:paulin03},
SLOTT-AGAPE \cite{ref:novati02,ref:novati03}, WeCAPP \cite{ref:riffeser03},
MEGA \cite{ref:mega04}, NainiTal \cite{ref:joshi05}.

Beyond the detection of viable microlensing candidates,
in order to draw conclusions on the physical issue of the
nature of the lenses, and therefore on the halo content in form of MACHOs,
one must develops models able to predict the expected signal.
With respect to the Magellanic Clouds searches a few important differences
arise. First, as a direct consequence of the unresolved-source issue,
for M31 experiments one has to model the luminosity functions of the sources.
A related issue is that the intrinsic M31 surface brightness shows a strong gradient
moving towards the galaxy centre, and this introduces a spatial dependent
noise level one has to take into account in order to correctly
predict the level of amplification needed, for a given source magnitude,
to give rise to a detectable microlensing event \cite{ref:kerins01}.

The second important difference is the ratio of expected self lensing
versus MACHO lensing, that in the case of observations towards 
M31 is much larger than for observations towards the LMC.
This is  a consequence of the fact that the M31 luminous
components are much more massive than the LMC one's.
The exact figure depends
on the observed field of view, the MACHO mass
and halo fraction (and also on the not-so-well-known self-lensing contribution). 
However, comparing for instance the MACHO and the POINT-AGAPE analyses
(to be discussed below) for full halos of $\sim~0.5~\mathrm{M}_\odot$ 
this ratio turns out to be \emph{larger}, in the M31 case, by about one order of magnitude.
As for the search of the MACHO lensing signal, this expected 
self-lensing signal constitutes therefore an unwanted background one must be able
to deal with and eventually to get rid of. On the other hand, a relatively large
self-lensing signal is important as it allows one to study the characteristics 
of the M31 stellar populations.

A complete analysis of the microlensing, observed and expected, signal
has been performed by the POINT-AGAPE \cite{ref:novati05} 
and the MEGA \cite{ref:mega06} collaborations. The two groups shared the same set of data,
taken at the 2.5m INT telescope over a period of 4 years (1999-2002),
but carried out completely independent analyses. As for the data analysis, 
in particular, POINT-AGAPE used the so-called ``pixel-photometry'' 
\cite{ref:agape97}, while MEGA used the DIA (difference image analysis) 
photometry \cite{ref:tomaney96}. Furthermore, the two groups followed
different strategies as for the determination of the efficiency
of their analysis pipeline, this step being the fundamental 
link between the theoretical predictions and the results
of the data analysis.

\begin{figure}
\includegraphics[width=9cm]{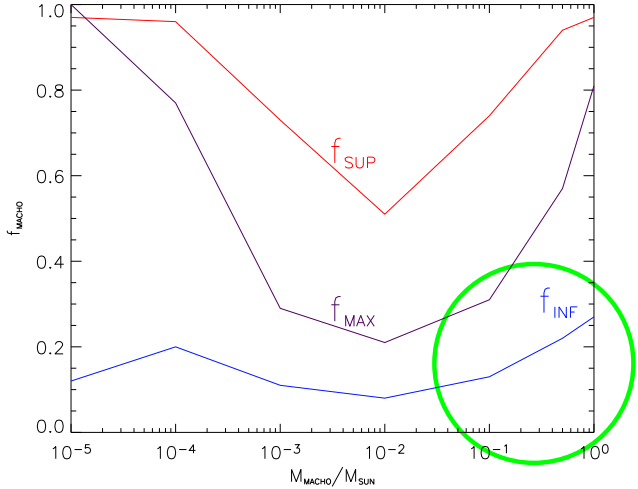}    
\caption{The POINT-AGAPE results: 
Most probable values, upper and lower 95\% CL limit for the halo fraction as
a function of the MACHO mass. (Figure adapted from \cite{ref:novati05}.)}
\label{fig:pa-res}
\end{figure}

The conclusions out of these two experiments turned out to be
in disagreement.  POINT-AGAPE claims for an evidence of a MACHO
contribution to Galactic halos, whereas MEGA finds his detected
signal to be compatible with self lensing.

POINT-AGAPE \cite{ref:novati05} restricted his search to bright microlensing events,
and reported the detection of 6 microlensing candidates, for
which the possible variable stars contamination was throughly discussed
and eventually discarded. Among these 6, one was found to be located
right along the line of sight to M32, a M31 satellite galaxy,
and attributed to an intergalactic M31-M32 microlensing event \cite{ref:paulin02}
and therefore excluded from the analysis of the halo content in form of MACHOs.
Through a Monte Carlo analysis of the expected signal it was then
shown that the expected self-lensing signal, for viable M31 luminous models,
was at most $\sim~1$ event, to be compared with $\sim~7$ events expected
for  full halos (both the Galactic and the M31) of $0.5~\mathrm{M}_\odot$ MACHOs.
Taking into account the spatial distribution of the observed events,
in that respect it turned out to be particularly relevant the position
of 1 event, located rather far away from the M31 centre (Figure~\ref{fig:int-fields}),
POINT-AGAPE  concluded claiming that ``[\ldots] the observed signal 
is much larger than expected 
from self lensing alone and we conclude, at the 95\% confidence level, that
at least 20\% of the halo mass in the direction of M31 must be in the form of MACHOs
if their average mass lies in the range $0.5-1~\mathrm{M}_\odot$ [\ldots]'' 
(Figure~\ref{fig:pa-res}).

MEGA \cite{ref:mega06} identified 14 microlensing candidates reaching,
however, an altogether different conclusion : ``[\ldots] the observed 
event rate is consistent with the rate predicted for self-lensing -
a MACHO halo fraction of 30\% or higher can be ruled out at the 95\% 
confidence level [\ldots]''.

These contradictory results give rises to a through debate.
One of the main point of disagreement is, in fact, the prediction
of the expected self-lensing signal and its characterisation with
respect to MACHO lensing (e.g. \cite{ref:kerins01,ref:baltz05,ref:riffeser06}).
The results of the MEGA experiment have been further analysed
in \cite{ref:ingrosso06,ref:ingrosso07}, where in particular
the spatial and the duration distributions of the observed events
has been considered, with the conclusion that self lensing
can not explain all the reported microlensing candidates.

\section{Conclusions}

Microlensing searches towards the Magellanic Clouds and the Andromeda galaxy, M31,
have given, in recent years, first exciting though somewhat contradictory conclusions.
The detection of microlensing events has been reported, however 
their interpretation with respect to the halo dark matter issue
is still open to debate. Along both line of sights ``evidence'' for a MACHO signal
as well as null results have been reported. (We may stress that, in the framework
of galaxy formation theory, even a relatively ``small'' halo fraction 
contribution in form of MACHOs,
say at the 10\%-20\% level, may turn out to be  relevant). New observational 
campaigns are currently under way to further address
this interesting issue. The SuperMACHO collaboration \cite{ref:rest05} is
observing the LMC with a much larger field of view than previous campaigns.
Towards M31, the Angrstom \cite{ref:kerins06} 
and the PLAN \cite{ref:loiano07} campaigns are currently underway.
Both, in different ways, aims rather to the central M31 region,
so to properly characterise the self-lensing signal. Furthermore,
as opposed to previous analyses, there is an effort
to get to a more suitable sampling of the observational data 
so to allow a better reconstruction of the microlensing event parameters,
in particular of the Einstein time.

\acknowledgments
It is a pleasure to thank the organisers
of this I Italian-Pakistan Workshop
of Relativistic Astrophysics for the warm 
ambience we enjoyed through this interesting
meeting. Thanks to Jean Kaplan for carefully
reading the manuscript.

\end{document}